\documentclass[a4paper,twocolumn]{article}

\usepackage{amsmath,amsfonts,amssymb}
\usepackage{amsthm}
\usepackage{enumerate}
\usepackage{bbm}
\usepackage{graphicx}
\usepackage{lscape,epsf}
\usepackage[vcentermath]{youngtab}
\usepackage{rotating}
\usepackage{hyperref}

\AtBeginDocument{\addtolength{\columnsep}{8pt}}
\AtBeginDocument{\addtolength{\textwidth}{10pt}}
\def\ds{\displaystyle}
\def\bea{\begin{array}{c}}
\def\ea{\end{array}}
\def\be{\begin{equation}\bea\ds}
\def\ee{\ea\end{equation}}
\def\bee{\begin{equation}\begin{array}{rcl}\ds}
\def\eee{\end{array}\end{equation}}

\def\nn{\nonumber}

\title{\bf Topological transport from a black hole}

\author{Dmitry Melnikov}

\date{}

\begin{document}



\twocolumn[
  \begin{@twocolumnfalse}
  \hfill{ITEP-TH-10/17}
    \maketitle
    \vspace{-0.5cm}
    \begin{center}
    {\textit{International Institute of Physics, Federal University of Rio Grande do Norte, \\ Campus Universit\'ario - Lagoa Nova,   Natal, RN 59078-970, Brazil}\\ \vspace{6pt}
\textit{Institute for Theoretical and Experimental Physics, \\B.~Cheremushkinskaya 25, Moscow 117218, Russia}\\ \vspace{6pt}}
    \end{center}
    \vspace{0.2cm}
    \begin{abstract}
In this paper the low temperature zero-frequency transport in a $2+1$ - dimensional theory dual to a dyonic black hole is discussed. It is shown that transport exhibits topological features: the transverse electric and heat conductivities satisfy the Wiedemann-Franz law of free electrons; the direct heat conductivity is measured in units of the central charge of CFT$_{2+1}$, while the direct electric conductivity vanishes; the thermoelectric conductivity is non-zero at vanishing temperature, while the $O(T)$ behaviour, controlled by the Mott relation, is subleading. Provided that the entropy of the black hole, and the dual system, is non-vanishing at $T=0$, the observations indicate that the dyonic black hole describes a $\hbar\to 0$ limit of a highly degenerate topological state, in which the black hole charge measures the density of excited non-abelian quasiparticles.
\end{abstract}
\vspace*{0.5cm}
  \end{@twocolumnfalse}
  ]

\paragraph{Introduction.} AdS/CFT is a powerful tool to approach a certain class of strongly coupled quantum systems. The method is based on a conjectured duality between string theory in anti-de Sitter (AdS) space and conformal theory (CFT) on the boundary of AdS~\cite{AdS/CFT}. When the string theory is in its low-energy weak-coupling limit of classical gravity the dual CFT is in a quantum strongly coupled phase. Henceforth we refer to this as the \emph{holographic} limit.

It turns out that the strong coupling regime of the CFT probed  by the duality is a peculiar one. In particular, it does not apply directly to strong interactions in particle physics, as originally expected, since it addresses the regime of extreme number of internal degrees of freedom (color) and extreme values of coupling constant. It was consequently proposed that a more natural domain of applicability of AdS/CFT belongs to condensed matter physics. An appropriate review can be found in Ref.~\cite{AdS/CMT}.

Following the idea of the proposal we would like to revisit the view of AdS/CFT on transport in $2+1$ - dimensional systems with finite charge density and magnetic field. The focus in this paper will be on the low-temperature transport. Based on the analysis of transport properties we claim that the simplest $3+1$ - dimensional dual gravity description of such a system predominantly reflects its topological features.

The most interesting observation that we will present here is that $3+1$/$2+1$ - dimensional ``holographic" duality is consistent with the $2+1$/$1+1$ bulk-to-boundary correspondence in well-known topological setups, such as quantum Hall effect (QHE). Most transparently, the heat conductivities, as computed by the gravity model, exhibit a typical behavior, consistent with CFT models of $1+1$ - dimensional edge modes in QHE. The low temperature results thus indicate exact systems, where predictions of AdS/CFT could be tested, even experimentally. Some experimental challenges are outlined in the conclusion to this paper. In particular, holography instructs us to work in a ``classical'' regime of degenerate topological states of matter.

The gravity system that has the above features is provided by an electrically and magnetically charged (dyonic) black hole~\cite{Hartnoll:2007ai}.

\paragraph{In dyonic black holes} transport was originally discussed by Hartnoll and Kovtun~\cite{Hartnoll:2007ai}, and later by seminal papers~\cite{Hartnoll:2007ih} and~\cite{Herzog} of Hartnoll \emph{et al}, which demonstrated an impressive consistence of holographic approach with a more conventional hydrodynamical one. The gravity side of the story is provided by a $3+1$ - dimensional Einstein-Maxwell theory with a negative cosmological constant. This theory has a solution corresponding to an asymptotically AdS black hole metric coupled to electric and magnetic fields. The latter fields are parallel to the fourth ``radial" AdS coordinate $z$, so that at the asymptotic boundary $z\to 0$, where the expected $2+1$ - dimensional dual theory lives, the electric field turns into a two-dimensional surface charge density $\rho$, while the magnetic field $B$ becomes a transverse magnetic flux.

Following the holographic prescription one can compute equilibrium thermodynamics of the dual system as well as response to external perturbations. Gravity calculation appears as powerful as the hydrodynamical one, yet it is more complete and provides information on the equation of state. Summarizing the zero frequency results from Refs.~\cite{Hartnoll:2007ai,Hartnoll:2007ih} on transport coefficients, classical gravity calculation in the dyonic black hole background expresses electric, thermal and mixed conductivities in terms of the thermodynamical quantities. While the result for electric conductivity does not seem to be very illuminating, namely, non-vanishing is only the transverse Hall conductivity, which is expressed as $\sigma_H=\rho/B$, the thermal conductivity is given by a less trivial expression:
\begin{eqnarray}
\kappa_{xx} \ = \ \kappa_{yy} & = & \frac{a s^2 T}{\rho^2+a^2 B^2}\,, \nn \\
\label{kperp}
\kappa_{xy} \ = \ -\,\kappa_{yx} & = & \frac{ \rho s^2 T}{B(\rho^2+a^2 B^2)} \,.
\end{eqnarray}
Here $T$ and $s$ are the temperature and entropy density of the thermodynamic system described by the black hole. The quantity
\be
\label{centralcharge}
a \ = \ \frac{L^2}{4G} \,,
\ee
comes from the gravitational/geometric parameters: $G$ is the four-dimensional Newton's constant and $L$ is the curvature radius of the AdS space. One can establish the precise meaning of $a$ on the dual side, if the black hole is embedded in a full string theory setup. In~\cite{Herzog:2007ij} it is identified as $\sqrt{2}N^{3/2}/6\pi$ in terms of a dual superconformal gauge theory with $SU(N)$ gauge group. More generally it is a parameter that characterizes a number of degrees of freedom of the dual CFT.

\paragraph{In the low temperature limit} thermodynamics of the dyonic black hole and  Eq.~(\ref{kperp}) yield the following result for the thermal conductivities
\begin{eqnarray}
\kappa_{xx}  & = &  \frac{\pi^2}{3}\,a T + O(T^2)\,, \nn \\
\label{kperp2}
\kappa_{xy} & = & \frac{\pi^2}{3}\,\sigma_H T + O(T^2)\,.
\end{eqnarray}

The numerical coefficient $\pi^2/3$ that appears in the expressions for the conductivities is the conventional quantum of thermal conductivity ``quantized" in units of parameters $a$ and $\sigma_H$. In particular, it was appreciated in Ref.~\cite{Melnikov:2012tb} that the ratio of transverse thermal and electric conductivities,
\be
\frac{\kappa_{xy}}{\sigma_{xy}} \ = \ \frac{\pi^2}{3}\,T\,,
\ee
yields the Wiedemann-Franz law for classical metals.

The expression for $a$ in terms of gravity parameters together with the form it appears in Eq.~(\ref{kperp2}) implies that we should identify $a$ with a central charge of the dual theory. In three-dimensional gravity in AdS space their exists a similar relation derived by Brown and Henneaux~\cite{Brown:1986nw}. Specifically, the boundary degrees of freedom of $AdS_3$ are governed by a $1+1$ - dimensional CFT with central charge $c=3L^{(3D)}/2G^{(3D)}$. Expression~(\ref{centralcharge}) for $a$ is a particular $D=2+1$ form of the central charge in odd dimensions conjectured by Myers and Sinha~\cite{Myers:2010xs} in the analysis of the universal contribution to entanglement entropy. Eq.~(\ref{kperp2}) provides further evidence to this conjecture.

\paragraph{In a topological state,} such as one in QHE, transport occurs at edges of the system. In the presence of edges the effective Chern-Simons theory of QHE requires massless boundary degrees of freedom to preserve gauge invariance. The theory of edge modes is a CFT, whose central charge is connected with the filling fraction of the QHE state~\cite{Wen}. CFT allows to compute the transverse Leduc-Righi (LR) conductivity
\be
\kappa_{xy} \ = \ \frac{\pi^2}{3}\,\nu_Q T\,,
\ee
where $\nu_Q$ is a sum over edge mode channels. In integer QHE, see \emph{e.g.}~\cite{Kane:1997fda},  $\nu_Q=\sigma_H$, and the holographic formula for the LR conductivity appears consistent. The difference between $\nu_Q$ and $\sigma_H$ appears when channels with different quasiparticle charges and, consequently, different $\sigma_H$ are present. This is not captured by the naive holographic picture.

In a 1998 paper~\cite{Read:1999ch} Green and Read proposed to derive the quantization of $\kappa_{xy}$, coupling energy current to external metric fluctuations. Similarly to the electric potential, metric would be controlled by a gravitational Chern-Simons theory. As we know, gravity has a Chern-Simons description in three dimensions. The approach suggested by Green and Read is what is now  ``routinely" applied in AdS/CFT.

In 3D the above result for the LR conductivity is easy to obtain. In $AdS_3$ an easy calculation yields
\be
\label{AdS3}
\kappa \ =\ \frac{\pi}{6} \,c T\,.
\ee
in terms of the Brown-Henneaux central charge $c$. Note that $2+1$ - dimensional gravity describes a $1+1$-dimensional system, which is the edge of the QHE bar. Thus $\kappa =\kappa_{xy}$ and $\sigma_H = c/2\pi$. Meanwhile the direct conductivity $\kappa_{xx}$ is defined in terms of a $2+1$ - dimensional central charge $a$.

The agreement between Eqs.~(\ref{kperp2}) and~(\ref{AdS3}) is quite interesting. We remind that while in $AdS_3$ this result can be easily obtained using conformal symmetry, in $AdS_4$ it becomes a much less trivial calculation, \emph{e.g.}~\cite{Hartnoll:2007ai}. We believe that the reason for the agreement lies in the fact that the $T\to 0$ result is topological.

\paragraph{In experiment} more accessible are the thermoelectric coefficients, which characterize the current or voltage response to an applied temperature difference. First, from formulae in~\cite{Hartnoll:2007ai,Hartnoll:2007ih} one finds the low temperature expansion of the thermoelectric conductivity:
\begin{eqnarray}
\alpha_{xx} & = & 0\,, \nn \\
\alpha_{xy} & = & - \alpha_{yx} \ = \ \frac{\pi}{\sqrt{3}}\sqrt{\sigma_H^2+a^2} + O(T)\,.
\end{eqnarray}
It appears that the off-diagonal part of $\alpha$ is a square root of the sum of $\kappa_{xx}^2$ and $\kappa_{xy}^2$ divided by the temperature.

The thermoelectric power (TEP) $S$ can be found from the matrix formula $S=-\sigma^{-1}\cdot\alpha$. We conclude that
\be
S_{xx}\ =\ -\,\frac{s}{\rho} \ = - \ \frac{\pi}{\sqrt{3}}\frac{\sqrt{\sigma_H^2+a^2}}{\sigma_H} + O(T)\,.
\ee
while the transverse components, measuring the Nernst response, vanish.

As in Boltzmann's theory the TEP is entropy carried by unit charge. The fact that it does not vanish for zero temperature is related to non-vanishing entropy at $T=0$. In other words, the ground state of the black hole is degenerate, which happens in topological states of matter. The leading $T=0$ coefficient thus computes the topological TEP.

It is interesting to compute the next order contribution to the (Seebeck) coefficient $S$. Using the black hole equation of state it can readily be presented in the form
\be
\label{Mott}
S_{xx} \ = \  S_{xx}^{\rm (top)} - \frac{\pi^2}{3}\,T\left(\rho^{-1}\frac{d\rho}{d\mu}\right) + O(T^2).
\ee
In the latter term one recognizes the Mott relation~\cite{Mott}.

\begin{figure}[t]
 \centering
 \includegraphics[width=\linewidth]{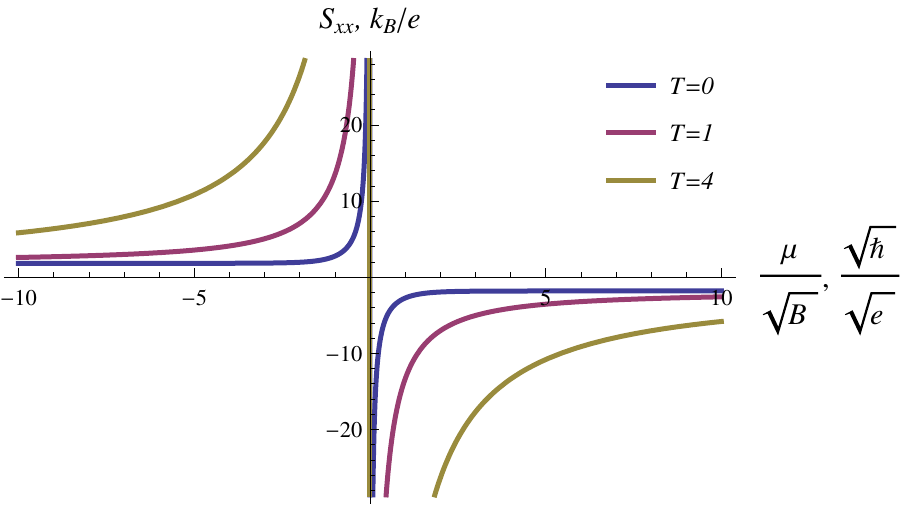}
 \caption{(color online) Seebeck coefficient as a function of the chemical potential (applied gate voltage). Different curves show the dependence at different temperatures, which is measured in units of $\mu/\sqrt{B}$.}
 \label{fig:tep}
\end{figure}

In figure~\ref{fig:tep} we show a plot of the coefficient $S_{xx}$ as a function of chemical potential and temperature. This plot can be compared with experimental data. In real QHE $S_{xx}$ would exhibit oscillations, between zero value (at a Hall plateau) and some maximum value (transition between plateaux), which is inversely proportional to the Landau level number. The connection between experimental behavior and the one on figure~\ref{fig:tep} will become clear in the following discussion.

Another application of the transport information is an estimate for the dimensionless figure of merit in holographic topological transport
\be
ZT\ = \ \sigma\cdot S^2\cdot \kappa^{-1}T \ =\ \left(\begin{array}{cc}
                     1 & \frac{a}{\sigma_H}\\
                     -\,\frac{a}{\sigma_H} & 1
                    \end{array}
\right). 
\ee
Here we define $ZT$ as a matrix, via multiplication of transport matrices.

\paragraph{In charged black holes} it is a common practice to impose the condition $A_t=0$ for the bulk (Maxwell) gauge field at the horizon. This ensures regularity of thermodynamic potentials, but also fixes the thermodynamical relation between the chemical potential and charge density. In the case of the dyonic black hole~\cite{Hartnoll:2007ai} the relation reads
\be
\label{thermorelation}
\rho \ = \ a\,\frac{\mu}{z_h}\,,
\ee
where $z_h$ is the horizon radius of the black hole in terms of the $AdS_4$ radial coordinate $z$~\cite{radius}, fixed by the relation \be
\label{horizon}
z_h^2\mu^2+z_h^4B^2+4\pi z_h T =3\,.
\ee Together, Eqs.~(\ref{thermorelation}) and~(\ref{horizon}) constitute the equation of state of the dual system. It is not hard to verify that Mott relation~(\ref{Mott}) was derived assuming the above relation.

We remind that the horizon radius is a geometric scale, which can be holographically associated with a physical energy scale in the dual theory. In black holes with no charge this scale is temperature. Heuristically, all physics characterized by an energy scale below the temperature scale gets swallowed by the black hole.

We are interested in the regime $T\to 0$ of the charged black hole. In this limit the black hole still has a finite radius. The geometric scale can now be associated with the chemical potential and/or the Landau level filling fraction as instructed by Eq.~(\ref{horizon}).

To sum up the observations, the system seems to be in a QHE-like state since the direct electric conductivity vanishes. One the other hand, the filling fraction is not quantized. Moreover, charge density $\rho$ depends on both the chemical potential and magnetic field. In the same time one observes that the zero temperature state of the system has a non-vanishing entropy.

Such a system has a simple interpretation. It is a holographic limit of a topological state. The charge density $\rho$ is not the density of electrons filling the Landau levels, but rather the density of quasiparticle excitations. The density of the quasiparticles scales as the central charge $a$. It must be large in the limit $a\to\infty$. This is similar to the classical limit $\hbar\to 0$, where one cannot distinguish individual plateux of conductivity and the latter appear as continuous rather than quantized.

A topological state with a large number of non-abelian quasiparticles can be highly degenerate. This is accounted by the entropy, which is also proportional to the central charge. Consequently, topological nature of this quasiparticle system is reflected in the behavior of the transport coefficients at low temperatures.

\paragraph{In conclusion} we summarize our observations about low-temperature, zero-frequency transport predicted by the dyonic black hole. The electric and heat conductivities have the following scaling
\begin{eqnarray}
 \sigma & = & \left(\begin{array}{cc}
                     0 & \sigma_H\\
                     -\sigma_H & 0
                    \end{array}
\right),\nn \\
\kappa & = & \frac{\pi^2}{3}\left(\begin{array}{cc}
                     a & \sigma_H\\
                     -\sigma_H & a
                    \end{array}
\right)T\,,
\end{eqnarray}
where $a$ is the central charge of the underlying CFT, large in the holographic limit. In this limit charge density scales as $O(a)$, which is equivalent to the classical limit $\hbar\to 0$ in $\sigma_H=\rho/B$, where it becomes continuous rather than quantized. The quantization should be recovered at small values of the chemical potential, $\mu\sim1/a$, where $\sigma_H=O(1)$ and $\delta\sigma_H/\delta\mu\to \infty$.

The thermoelectric conductivity is transverse, $\alpha_{xy}=\sqrt{\kappa_{xx}^2+\kappa_{xy}^2}/T$. At low temperature it is independent from $T$. So is the Seebeck constant (TEP), which in the limit of large $a$, but finite $\sigma_H$, scales as
\be
S \ = \ - \frac{\pi a}{\sqrt{3}\sigma_H}\,.
\ee
This result can be used as a reference for experimental values of heights of the maxima of $S_{xx}$ in topological phases. Indeed, in the $a\to\infty$ limit, what the plot on figure~\ref{fig:tep} must be showing is the envelope curve of the oscillating experimental function. The characteristic figure of merit of the topological system shows the same scaling $a/\sigma_H$.

We remind that the results cited in Refs.~\cite{Hartnoll:2007ih} and~\cite{Herzog} apply for any values of $T$, and further, any frequency $\omega$. It would be interesting to analyze the experimental consequences of those results also departing from the low-temperature regime. We have shown that the subleading temperature behavior of TEP is given by the Mott relation, which is often a good description of experimental results. It would be interesting if the leading behavior could be tested in a degenerate topological state. For example, a non-vanishing TEP at zero temperature is consistent with a non-abelian nature of the quasiparticles, \emph{cf.} Ref.~\cite{Halperin}.

A more challenging experimental task is to test the subleading behaviour of heat conductivities. Expanding Eqs.~(\ref{kperp}) in small $T$ we find that the ratio of the $O(T^2)$ coefficients can be expressed in terms of topological data as
 \be
\frac{\kappa_{xx}^{(2)}}{\kappa_{xy}^{(2)}} \ = \ \frac{2a(2a^2+\sigma_H^2)}{\sigma_H(5a^2+3\sigma_H^2)}\,.
 \ee
At large $a$, but finite $\sigma_H$, this ratio tends to $4a/5\sigma_H$.

An interesting theoretical question is what this topological behavior means for the $AdS_4$ Einstein-Maxwell theory itself. Perhaps, by connecting it to an appropriate ``Chern-Simons'' theory on the boundary, as in~\cite{Herzog:2007ij}, the theory could prove completely solvable.

\paragraph{Acknowledgements} The author benefited from enlightening discussions with A.~Abanov, D.~Giataganas, D.~Khveshchenko, A.~Mironov, A.~Morozov, F.~Novaes, R.~Pereira and P.~Wiegmann as well as from many interesting conferences held at the IIP in Natal. He would also like to thank S.~Klevtsov, N.~Toumbas and the hospitality of the University of Cologne and the University of Cyprus, where a part of this work was done. The work was supported by the Russian Science Foundation Grant No.~16-12-10344.


\begin{thebibliography}{99}
\bibitem{radius} It is a bit sloppy to call $z_h$ the black hole radius. A more proper radius is given by $L^2/z_h$: when black hole is ``large" -- $z_h$ is small.
\end{thebibliography}
\end{document}